\documentclass[prl,aps,twocolumn,groupaddress]{revtex4-1}

\usepackage[centertags]{amsmath}
\usepackage{graphicx}
\usepackage{amsfonts}
\usepackage{epstopdf}
\usepackage{color}
\usepackage[normalem]{ulem}

\begin{document} 
\title{Nonconservative higher-order hydrodynamic modulation instability}
\author{O. Kimmoun$^{1,*}$, H. C. Hsu$^{2}$, B. Kibler$^{3}$ and A. Chabchoub$^{4,\dagger}$}

\affiliation{$^1$ Aix-Marseille University, CNRS, Centrale Marseille, IRPHE, Marseille, France}
\email{olivier.kimmoun@centrale-marseille.fr}
\affiliation{$^2$ Tainan Hydraulics Laboratory, National Cheng Kung University, Taiwan}
\affiliation{$^3$ Laboratoire Interdisciplinaire Carnot UMR 6303 CNRS / UBFC, Universit\'e Bourgogne Franche-Comt\'e, 21078 Dijon, France}
\affiliation{$^4$ Department of Mechanical Engineering, Aalto University, 02150 Espoo, Finland} 
\email{amin.chabchoub@aalto.fi} 

\begin{abstract}
The modulation instability (MI) is a universal mechanism that is responsible for the disintegration of weakly nonlinear narrow-banded wave fields and the emergence of localized extreme events in dispersive media. The instability dynamics is naturally triggered, when unstable energy side-bands located around the main energy peak are excited and then follow an exponential growth law. As a consequence of four wave mixing effect, these primary side-bands generate an infinite number of additional side-bands, forming a triangular side-band cascade. After saturation, it is expected that the system experiences a return to initial conditions followed by a spectral recurrence dynamics. Much complex nonlinear wave field motion is expected, when the secondary or successive side-band pair that are created are also located in the finite instability gain range around the main carrier frequency peak. This latter process is referred to as higher-order MI. We report a numerical and experimental study that confirm observation of higher-order MI dynamics in water waves. Furthermore, we show that the presence of weak dissipation may counter-intuitively enhance wave focusing in the second recurrent cycle of wave amplification. The interdisciplinary weakly nonlinear approach in addressing the evolution of unstable nonlinear waves dynamics may find significant resonance in other nonlinear dispersive media in physics, such as optics, solids, superfluids and plasma.  
\end{abstract}

\maketitle

One possible explanation for the formation of extreme wave events for instance in the ocean and nonlinear optical media is the modulation instability (MI) \cite{Bespalov,benjamin1967disintegration,biondini2016universal}. Understanding the wave dynamics of modulationally unstable waves is of major significance for the sake of accurate modeling and prediction of localized structures as well as of rogue waves in particular \cite{dudley2014instabilities,residori2016rogue}. The MI describes the disintegration of uni-directional and narrow-banded wave fields. Physically, the instability is driven, when side-bands that are located around the main carrier energy peak in a specific instability range are excited. The progressive focusing of the wave field is translated in spectral domain with an advancing formation of an infinite number of side-bands in form of a triangular cascade \cite{tulin1999laboratory,zakharov2009modulation}. 

One deterministic way to study the MI is by use of the nonlinear Schr\"odinger equation (NLSE) \cite{lighthill1965contributions,zakharov1968stability}. The latter weakly nonlinear evolution equation is indeed very useful in the study of the problem, in view of its integrability \cite{shabat1972exact}. In fact, the NLSE admits a family of exact solutions that model stationary, pulsating and modulationally unstable wave fields \cite{akhmediev1997solitons}. The standard model that describes the MI process are the family of Akhmediev breathers (ABs) \cite{akhmediev1985generation}. Indeed, for each unstable modulation frequency, or unstable side-band, one can assign an exact analytical AB expression to study the spatio-temporal evolution of the wave field. From an experimental perspective these universal type of solutions are very valuable, since the complex nonlinear physical processes can be controlled in time and space and adjusted to laboratory environments \cite{chabchoub2016hydrodynamic,kibler2016experiments}.  

It is also known that for non-ideal input conditions that the wave field experiences a focusing recurrence after the first growth and decay cycle of instability, also known as Fermi-Pasta-Ulam recurrence (FPU) \cite{fermi1955studies,tulin1999laboratory,dudley2009modulation,kimmoun2016modulation}. Interestingly, under particular conditions when the primary side-bands are shifted closer to the main frequency peak, the secondary or higher side-bands may also fall within the unstable frequency range. As consequence, it is expected that focused wave packets undergo a pulse splitting followed by much complex nonlinear wave interaction compared to a {\it standard} FPU recurrence dynamics. This process is referred to as higher-order modulation MI \cite{yuen1982nonlinear,erkintalo2011higher} and has been so far observed experimentally only in Kerr media \cite{erkintalo2011higher,hammani2011peregrine}. 

In this letter, we report the observation of higher-order MI on the water surface using the framework of ABs, taking into account weak dissipation that is present in our laboratory set-up and that impacts the nonlinear wave propagation motion. Experiments have been conducted in an unique and very large hydrodynamic wave facility, allowing the observation of a long-ranging  evolution of unstable wave dynamics. We also show that weak dissipation, usually and naturally present in laboratory environments, may counter-intuitively enhance the second recurrent focusing in a higher-order MI regime. The laboratory measurements, describing the higher-order dynamics, are in very good agreement with corresponding and nonconservative modified NLSE (MNLSE) \cite{dysthe1979note,trulsen2001spatial} simulations and justify the relevance of universal evolution equations in the study of nonlinear wave propagation in dispersive media. 

The dynamics of surface gravity waves in deep-water can be described by the framework of the NLSE \cite{osborne2010nonlinear}
\begin{equation} 
\operatorname{i}\left(\Psi_x+\dfrac{2k}{\omega}\Psi_t\right)-\dfrac{1}{g}\Psi_{tt}-k^3\left|\Psi\right|^2\Psi=0,
\label{NLS}
\end{equation} 
where $g$ denotes the gravitational acceleration and $k$ is the wavenumber of the narrow-banded as well as uni-directional wave field, that is connected to the wave frequency $\omega$ through the linear dispersion relation $k=\dfrac{\omega^2}{g}$. The NLSE is the simplest evolution equation that takes into account dispersion and nonlinearity of the wave dynamics. It can be also regarded as an universal evolution equation that describes wave dynamics in other field of physics, such as solids \cite{chong2014damped,grolet2016travelling}, optics \cite{dudley2014instabilities,walczak2015optical}, superfluid helium \cite{ganshin2008observation} and plasma \cite{bailung2011observation}. The particularity of the NLSE is its integrability. Indeed, it admits a number of stationary and pulsating localized envelope structures. One particular NLSE solution that describes the dynamics of MI of a Stokes wave of amplitude $a$ is the known as Akhmediev breather \cite{akhmediev1987exact}
\onecolumngrid 
\begin{eqnarray}
\Psi\left(X,T\right)=a\dfrac{\sqrt{2\mathfrak{a}}\cos\left(a\Omega T\right)+\left(1-4\mathfrak{a}\right)\cosh\left(2a^2RX\right)+\textnormal{i}R\sinh\left(2a^2RX\right)}{\sqrt{2\mathfrak{a}}\cos\left(a\Omega T\right)-\cosh\left(2a^2RX\right)}\exp\left(2a^2\textnormal{i}X\right).
\label{AB}
\end{eqnarray}
\twocolumngrid 
Here, $0<\mathfrak{a}<0.5$ denotes the breather parameter $R=\sqrt{8\mathfrak{a}\left(1-2\mathfrak{a}\right)}$ the growth and decay rate, $\Omega=2\sqrt{1-2\mathfrak{a}}$ the modulation frequency, while $X=\dfrac{k^3}{2}x$, $T=\sqrt{2}k^2\left(x-c_g t\right)$. Note that when $\mathfrak{a}\longrightarrow 0.5$, the modulation period becomes infinite, the growth becomes algebraic rather than exponential and the wave dynamics is then described by the universal Peregrine breather solution \cite{peregrine1983water,kibler2010peregrine,chabchoub2011rogue,tikan2017universal}. The AB-type wave motion has been observed in a wide-range of physical media and provides an ideal framework to control MI in space and time in laboratory environments \cite{dudley2009modulation,chabchoub2014hydrodynamics,kimmoun2016modulation}. In order to study numerically and / or experimentally the evolution dynamics of modulationally unstable Stokes waves, the initial input wave field of amplitude $a$ can be determined by 
\begin{eqnarray}
\Psi\left(x=x_0,t\right)=a\left[1+a_{\textnormal{mod}}\cos\left(\Omega t\right)\right]. 
\label{cosperturbation}
\end{eqnarray} 
To ensure intial AB dynamics as described in Eq. (\ref{AB}), the relation between modulation amplitude $a_{\textnormal{mod}}$ and the modulation frequency $\Omega$ should be as the following $\Omega=\sqrt{\dfrac{2R\mu \operatorname{i}}{a_{\textnormal{mod}}-\mu}}$, where $\mu$ is a real parameter \cite{dudley2009modulation,hammani2011spectral}. Even tough experiments in optics and hydrodynamics can be well-controlled, the presence of weak dissipation is inevitable. For water waves one possible source of dissipation is the viscosity. An effective model to take this into account is by adding a linear attenuation factor of wave envelope in the NLSE framework. Thus, for a given viscosity parameter $\nu$ the NLSE becomes \cite{dias2008theory}
\begin{equation} 
\operatorname{i}\left(\Psi_x+\dfrac{2k}{\omega}\Psi_t\right)-\dfrac{1}{g}\Psi_{tt}-k^3\left|\Psi\right|^2\Psi=-\operatorname{i}\mathfrak{D}\Psi,
\label{NLSD}
\end{equation} 
where $\mathfrak{D}=\dfrac{4k^3}{\omega}\nu$. It is also known that when ensuring a long propagation distance of the wave field, the MI undergoes a recurrent focusing \cite{tulin1999laboratory}. In this case, when only the primary side-band pair is within the unstable frequency range the unstable waves manifest a FPU-like growth-decay cycles \cite{dudley2009modulation}. When weak dissipation is at play, the cycle exhibits a specific shift, commuting crest and trough dynamics in each focusing cycle \cite{kimmoun2016modulation}. In the presence of strong dissipation, the modulation instability can be also completely annihilated \cite{segur2005stabilizing}. When the secondary or higher side-band pairs fall into the standard MI frequency range, complex dynamics of the wave field arises from splitting of the focusing localized structures and in some cases is followed by a nonlinear interaction of wave envelopes \cite{erkintalo2011higher}. 
\begin{figure}[h]
\begin{tabular}{cc}
\includegraphics[width=\columnwidth]{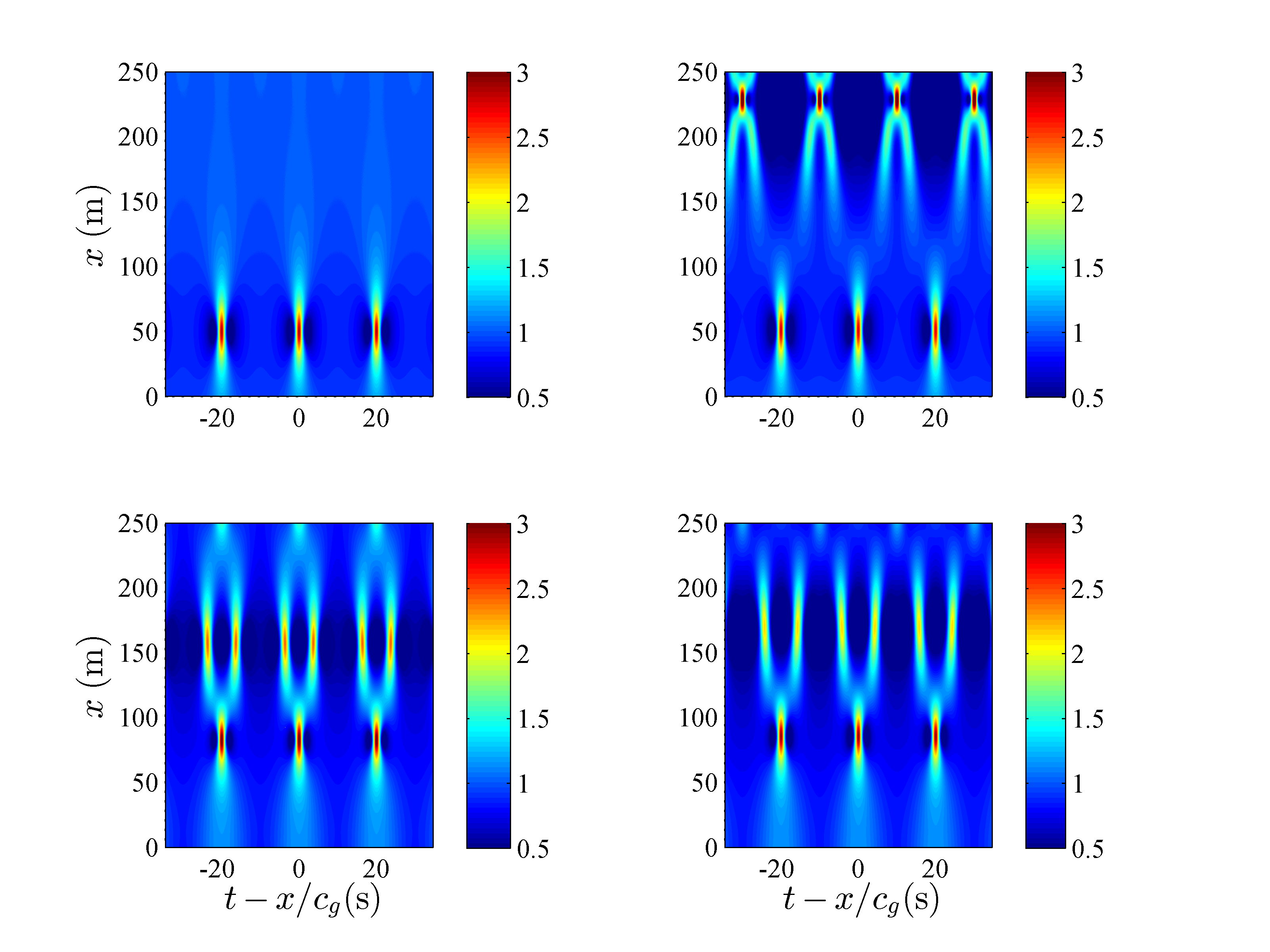}\\
\end{tabular}
\caption{(Color online) Top left: NLSE envelope evolution simulations of a conservative AB.  Top right: NLSE envelope evolution simulations of a dissipative AB with $\nu=1.2\cdot 10^{-5}$. Bottom left: NLSE simulations of a conservative envelope evolution with an approximated cosine modulation that fits the theoretical AB. Bottom right: NLSE simulations of a dissipative envelope evolution with an approximated cosine modulation that fits the theoretical AB with $\nu=1.2\cdot 10^{-5}$. The carrier parameters have been chosen to be $\varepsilon=ak=0.12$ with amplitude $a=0.03$ m at $x=-50$ m, the modulation frequency is determined by $\mathfrak{a} = 0.45$ while the envelope amplitudes have been normalized by the value of the amplitude $a$.} 
\label{fig1}
\end{figure}
Within the context of ABs, this can be achieved by shifting the modulation frequency towards $\Omega\longrightarrow 0$ or the breather parameter towards $\mathfrak{a}\longrightarrow 0.5$. More precisely, when $\mathfrak{a}\in]0.375;0.5[$. Since the NLSE solutions are very sensitive to noise, dissipation or small imperfections, it is never expected that the AB trajectory will converge back to the stationary Stokes or regular state \cite{kimmoun2016modulation}. 

Fig. \ref{fig1} shows the dimensional wave dynamics of an AB model Eq. (\ref{AB}) for $\mathfrak{a}=0.45$ as well as corresponding fitted periodic cosine envelope Eq. (\ref{cosperturbation}) in a conservative and dissipative context for carrier parameters $\varepsilon=ak=0.12$ and $a=0.03$ m. Clearly, a wave envelope patterns distinction can be noticed in higher-order MI regime, either in the conservative or dissipative as well as either within the AB framework or cosine envelope approach approximation. 

\begin{figure}[h]
\begin{tabular}{cc}
\includegraphics[width=.49\columnwidth]{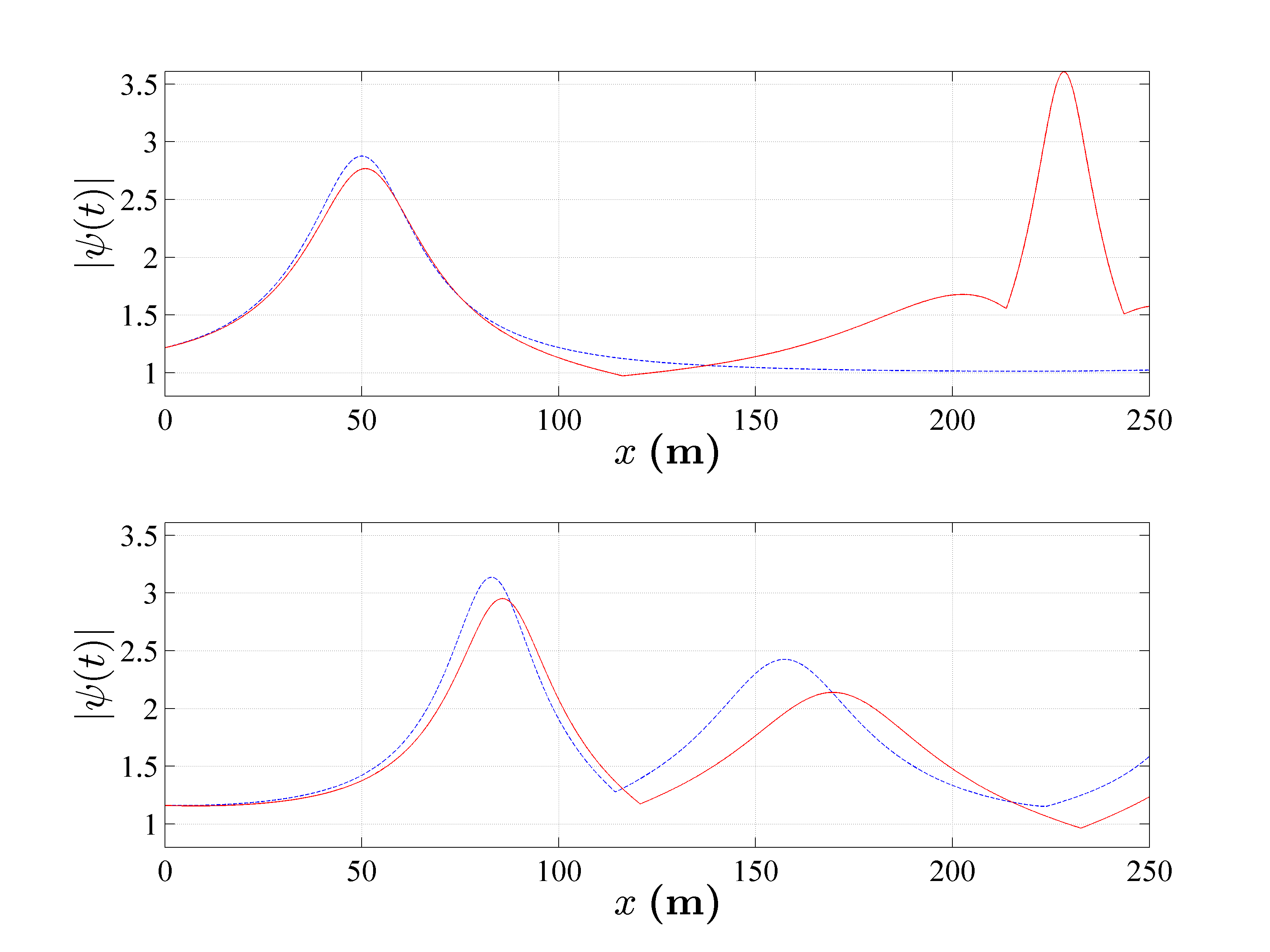}&
\includegraphics[width=.49\columnwidth]{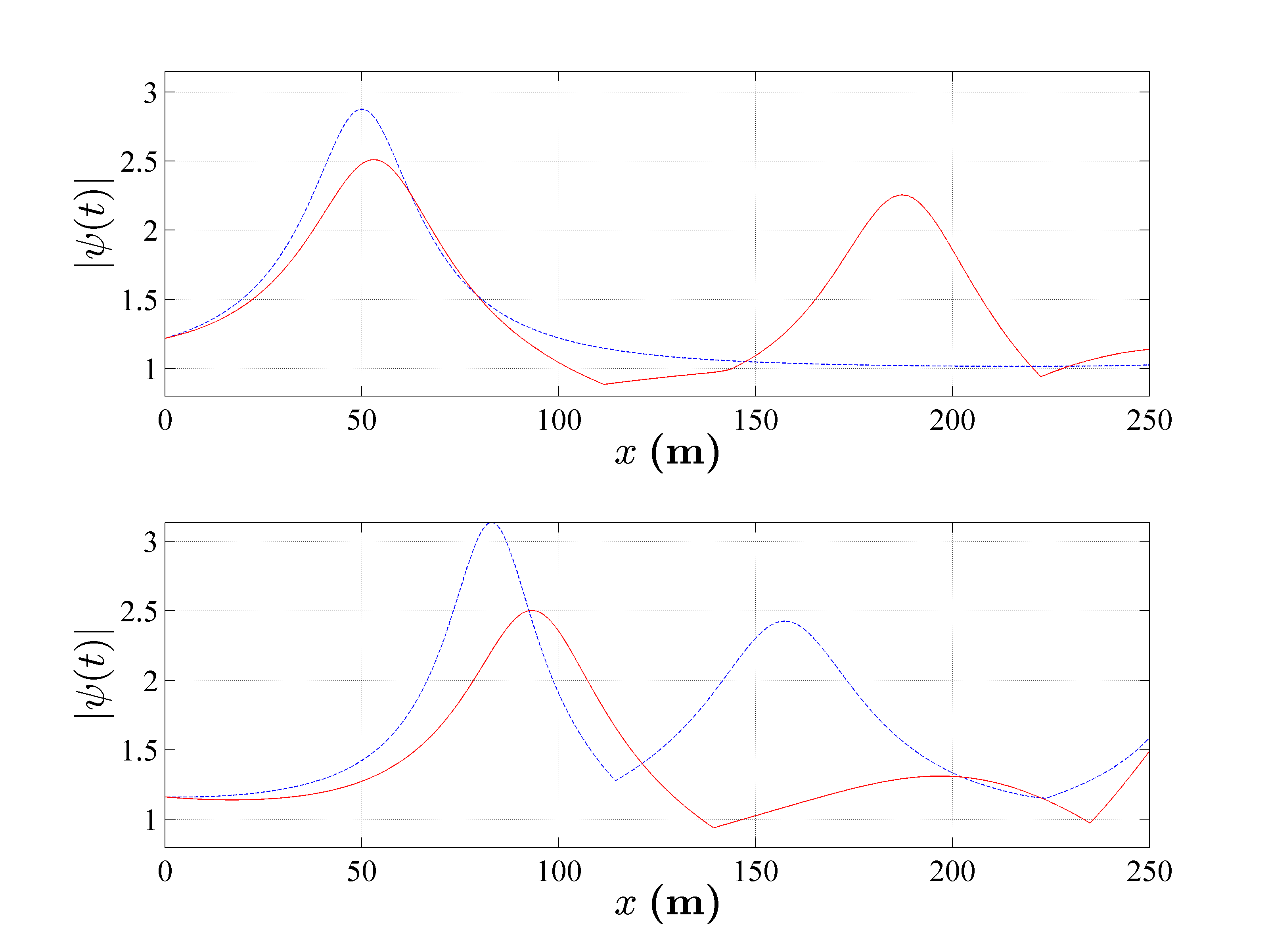}\\
\end{tabular}
\caption{(Color online) Maximal wave amplitude amplifications $\left|\psi\left(t\right)\right|$ in higher-order MI regime for carrier parameters $a=0.03$ m and $\varepsilon=0.12$ and $\mathfrak{a}=0.45$ over a propagation distance of $x=250$ m. Blue lines: Conservative dynamics. Red lines: Dissipative dynamics. Upper panels: AB dynamics. Lower panels: Corresponding cosine perturbation, as described by Eq. (\ref{cosperturbation}). Left panels: Dissipation rate is determined by $\nu=1.2\cdot 10^{-5}$. Right panels: Dissipation rate is determined by $\nu=4.1\cdot 10^{-5}$.
}
\label{fig2}
\end{figure}

We would like to briefly point out that the dissipation parameter allows to control the complex spatio-temporal arrangement (nonlinear superposition) of breather-type structures. In fact, for each modulation frequency in this regime, determined by $\mathfrak{a}\in]0.375;0.5[$ we can find a set of dissipation parameters $\mathfrak{D}$, or alternatively $\nu$, that engenders a collision of the {\it fissioned} ABs and as result a significant focusing is expected due to the nonlinear interaction of these. Fig. \ref{fig2} shows the normalized maximal amplitude amplifications $\psi=\dfrac{\Psi}{a}$ reached in the case of higher-order MI regime for an AB case and approximated cosine modulation in the conservative as well as dissipative framework. 
As expected, the first focusing is retarded when approximating the AB dynamics by a cosine modulation for $\mathfrak{a}=0.45$, as described by Eq. (\ref{cosperturbation}) \cite{dudley2009modulation}. More interestingly, when dissipation is at play for the chosen dissipation value that is determined by $\nu=1.2\cdot 10^{-5}$ the following second wave focusing of the initial AB envelope is significantly more amplified. Indeed, this second focusing is shown to be much higher than expected from standard AB or standard MI predictions. Namely, it is beyond three times the amplitude of the background \cite{osborne2010nonlinear,onorato2011triggering}. This type of dynamics at play for this case, as shown in the upper right and upper left panel of Figs. \ref{fig1} and \ref{fig2}, respectively, resembles the formalism and observations of AB collisions \cite{frisquet2013collision}. Since these significant amplifications do not occur when approximating the AB envelope by a cosine approach shows the importance of phase-shift dynamics in the dissipative process.  

\begin{figure}[h]
\centering
\includegraphics[width=\columnwidth]{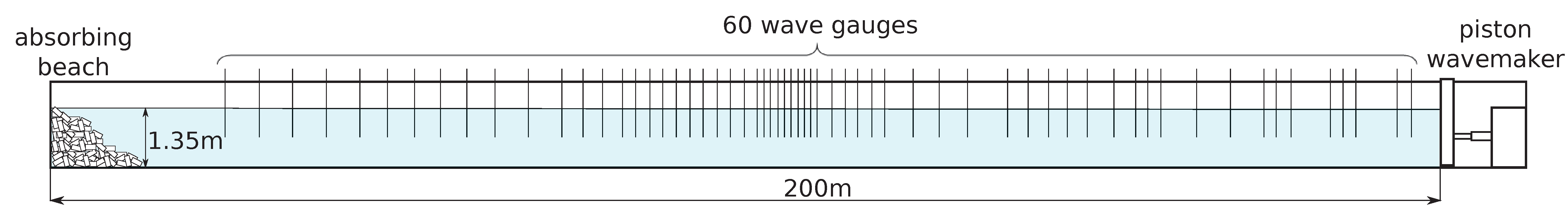}
\caption{(Color online) Schematic description of the water wave facility, installed at the Tainan Hydraulics Laboratory.}\label{fig3}
\end{figure}

Next, we describe the experimental as well as corresponding numerical investigation, related to higher-order MI wave dynamics on the water surface within the framework of AB envelope dynamics. Experiments have been conducted in a large water wave facility, installed at the Tainan Hydraulics laboratory. The facility has a length of 200 m with a constant water depth of 1.35 m while 60 capacitance wave gauges are installed along the flume. The small spacing between each wave probe allows a unique, precise as well as accurate wave field acquisition. This is a decisive fact, when aiming for the reconstruction of the wave's envelopes in order to compare these with weakly nonlinear NLSE-type predictions. Fig. \ref{fig3} shows the schematic illustration of the water wave flume as well as the placement of the wave gauges along the facility. 

In order to generate higher-order MI hydrodynamics we injected two different AB-type wave fields for $\mathfrak{a}=0.45$, thus, the case when the second and third side-band pairs are within the unstable range with exponential growth rate \cite{yuen1982nonlinear,osborne2010nonlinear,erkintalo2011higher}. The starting dynamics is initiated by taking into account small incipient envelope modulation, so that the first focusing occurs between 40 to 60 m from the wave maker. For the sake of validation of experimental results, the envelope of the measured water surface dynamics has been reconstructed, using the Hilbert transform \cite{osborne2010nonlinear} and then compared to NLSE and MNLSE simulations, including dissipation. The dissipative MNLSE formalism can be described by \cite{carter2016frequency} 
\begin{align} 
&\operatorname{i}\left(\Psi_x+\dfrac{2k}{\omega}\Psi_t\right)-\dfrac{1}{g}\Psi_{tt}-k^3\left|\Psi\right|^2\Psi=\nonumber\\
&\operatorname{i}\dfrac{k^3}{\omega}\left(6\left|\Psi\right|^2\Psi_t+2\Psi\left(\left|\Psi\right|^2\right)_t-2\operatorname{i}\Psi\mathcal{H}\left[\left(\left|\Psi\right|^2\right)_t\right]\right)+\nonumber\\
&\operatorname{i}\dfrac{k^3}{\omega}\left(-4\nu\Psi-20\operatorname{i}\dfrac{\nu}{\omega}\Psi_t\right).
\label{MNLSD}
\end{align} 
The linear dissipation rate $\mathfrak{D}$ has been determined in a prior experimental setting for a regular wave field with same corresponding wave parameters. Then, the parameter $\nu$ has been derived according to the relationship described in Eq. (\ref{NLSD}). The experimental results together with the numerical NLSE (\ref{NLSD}) and MNLSE (\ref{MNLSD}) predictions are shown in Fig. \ref{fig4}. 
{\onecolumngrid
\begin{center}
\begin{figure}[h]
\centering
\includegraphics[width=.99\textwidth]{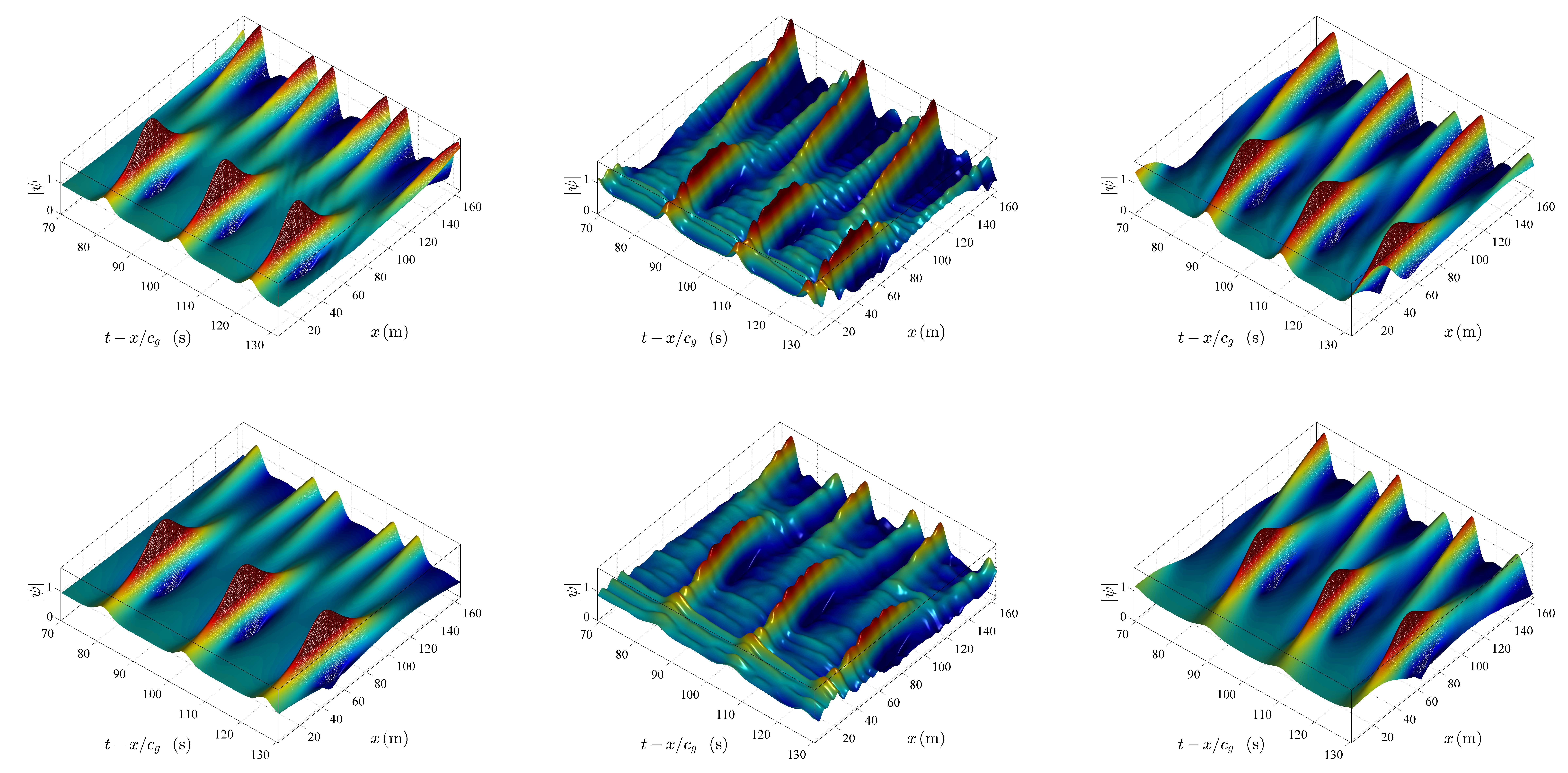} 
\caption{(Color online) From right to left: NLSE (\ref{NLSD}), experimental results and MNLSE (\ref{MNLSD}) simulations. Upper pannel: AB wave envelope evolution with parameter $\mathfrak{a}=0.45$ as well as carrier parameters $\varepsilon=0.10$ and $a=0.017$ m for $\nu=2.2\cdot 10^{-5}$. Bottom pannel: AB evolution with parameter $\mathfrak{a}=0.45$ as well as carrier parameters $\varepsilon=0.12$ and $a=0.03$ m for $\nu=4.1\cdot 10^{-5}$.}
\label{fig4}
\end{figure}
\begin{figure}[h]
\centering
\includegraphics[width=.99\textwidth]{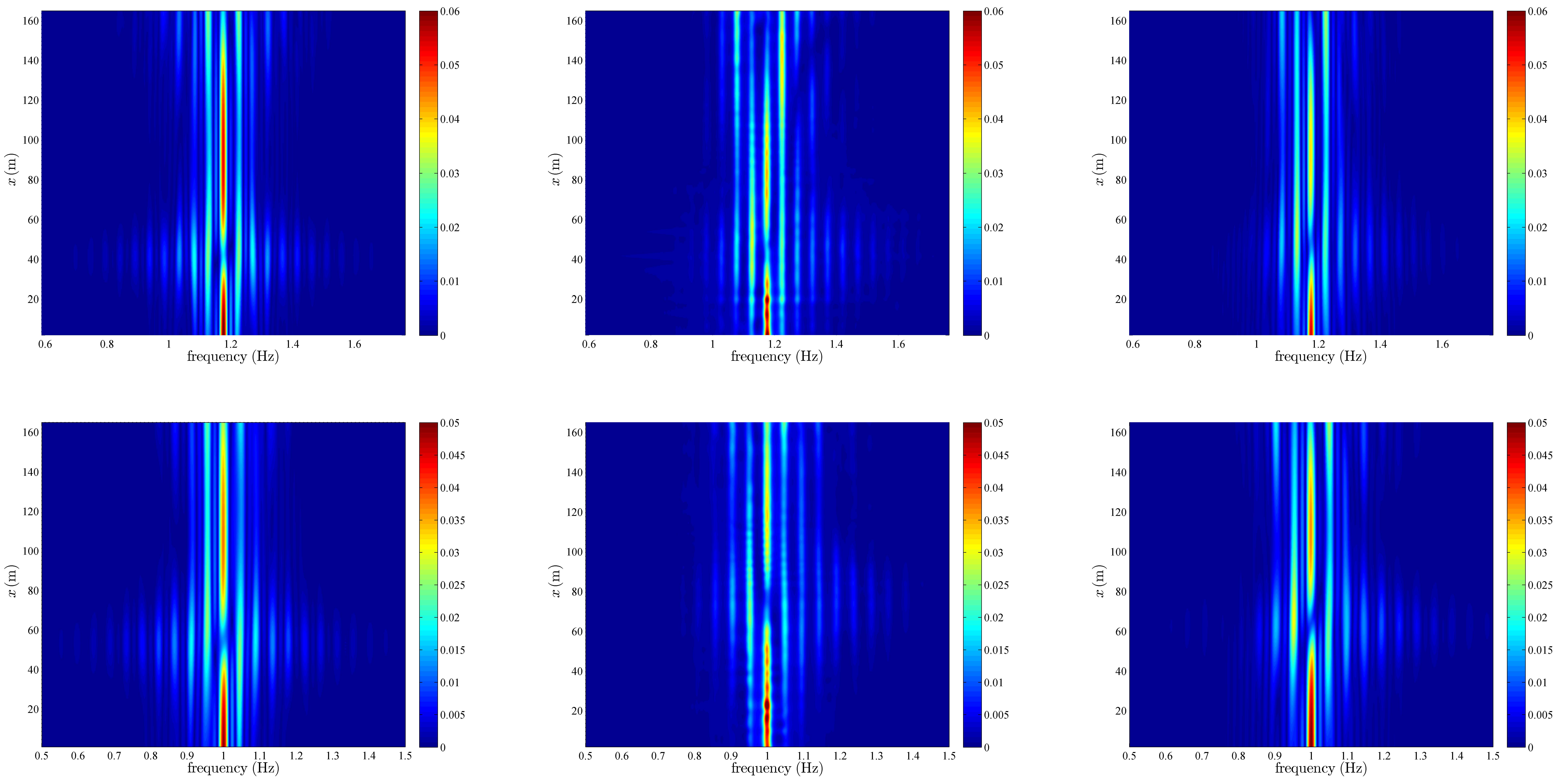}\\
\caption{(Color online) From right to left: NLSE (\ref{NLSD}), experimental and MNLSE (\ref{MNLSD}) spectral evolution of the wave field. Upper pannel: Spectral dynamics along the wave flume with AB parameter $\mathfrak{a}=0.45$ as well as carrier parameters $\varepsilon=0.10$ and $a=0.017$ m for $\nu=2.2\cdot 10^{-5}$. Bottom pannel: Spectral dynamics along the wave flume with AB parameter $\mathfrak{a}=0.45$ as well as carrier parameters $\varepsilon=0.12$ and $a=0.03$ m for $\nu=4.1\cdot 10^{-5}$.}
\label{fig5}
\end{figure}
\end{center}
\twocolumngrid}

Although the tank has a noticeable length of 200 m, we have not been able to observe the entire second focusing cycle that would reveal the corresponding complete dynamics of {\it fissioned} AB-type envelopes for the chosen wave and breather configuration parameters. However, we can clearly see the initialization of this focusing process, particularly, in the case shown in the upper panel of Fig. \ref{fig4}. The propagation distance required to observe the latter nonlinear dynamics can be reduced by increasing the value of the wave steepness. However, increasing the steepness engenders breaking of these steep focused AB breather-type waves as a result of significant amplification, thus, weakly nonlinear theories fail in describing such complicated wave dynamics \cite{babanin2011breaking}. Furthermore, due to the high wave amplitude amplifications reached for the chosen values of AB parameter $\mathfrak{a}$ as well as for the choice of carrier steepness, we can clearly notice and state that the MNLSE predictions are indeed more accurate than the NLSE model forecast. This can be notified in the asymmetry of wave envelope profiles in physical space. This is indeed well-captured in the MNLSE approach and can be explained by assessing the effects of higher-order dispersion and mean-flow \cite{dysthe1979note,trulsen2001spatial,lo1985numerical}. The latter asymmetry is obviously also translated to an asymmetry in the spectral Fourier space. We also emphasize that the dissipation in the model, that is determined by the experiment in the specific laboratory environment, does not allow the observation of a second wave amplification focusing that is higher compared to the first focusing cycle. The corresponding spectral evolutions are depicted in Fig. \ref{fig5}. Note that we just turn our attention on the spectral dynamics around the carrier peak frequency and excluded the dynamics of the higher-order Stokes harmonics (bound waves), that is, frequencies around $\dfrac{\omega}{\pi}$, $\dfrac{3\omega}{2\pi}$, $\dfrac{2\omega}{\pi}$ etc., in the spectral domain. These spectral evolutions clearly show the nonlinear complexity of the wave dynamics at play as well as the expected presence of strong asymmetry around the main carrier energy. In addition, we can perceive the initialization of the second wave focusing, characterized by the beginning of spectral broadening that is clearly annotated by the spectral broadening. This is another clear proof for the quantitative accuracy of the MNLSE approach when studying higher-order MI processes. This is also a first-time excellent comparison of long-term spectral evolution dynamics of MI in hydrodynamics. 

To conclude, we studied numerically and experimentally higher-order MI wave dynamics for surface gravity water waves in the presence of weak dissipation. The initial conditions for the experiments have been provided through the NLSE deterministic AB framework, complementing for instance experimental studies in optics in which the wave dynamics have been initiated from non-ideal breather input conditions. We discussed the possibility of counter-intuitive higher second wave focusing in a dissipative regime that is a result of AB-type envelope collision. The respective higher-order MI laboratory experiments reported are in very good agreement with numerical (dissipative) MNLSE simulations and confirm the applicability of weakly nonlinear models in the study of nonlinear water waves including extreme events, also when dissipation is at play. Furthermore, we showed that the dissipation parameter can be regarded as a new degree of freedom to control MI dynamics. We anticipate further studies in several nonlinear dispersive media with respect to higher-order MI, also to overcome the experimental limitations in hydrodynamics that are summarized in a short wave propagation distance and limited dissipation parameter range. Future work may be also devoted to the theoretical analysis of the effect of dissipation in the NLSE modeling \cite{baronio2014vector,wetzel2016experimental} as well as prediction of extreme waves \cite{cousins2015unsteady,randoux2016inverse} in various physical media governed by the NLSE-type equations. 

O.K. acknowledges support from the French-Taiwanese ORCHID Program of the Hubert Curien Partnership (PHC). H.C.H. acknowledges the grant support from MOST 104-2628-E-006-014-MY3, 105-2923-E-006-002-MY3 and 105-2611-I-006-301. B.K. acknowledges support from French project PIA2/ISITE-BFC. B.K. and A.C. are thankful for support from Burgundy Region (PARI Photcom). A.C. acknowledges Miguel Onorato, G\"oery Genty and John Dudley for fruitful discussions. 

\bibliography{Refs}

\end{document}